\documentclass[preprint,12pt]{elsarticle}

\usepackage[utf8]{inputenc} 

\usepackage[margin=3cm]{geometry} 
\usepackage{caption}
\usepackage{subcaption}
\usepackage{placeins}

\usepackage{amssymb}

\usepackage{hyperref}

\usepackage{listings}
\usepackage{color}

\newcommand{\Doxygen}{\textsf{Doxygen\/}} 
\newcommand{\Flowgen}{\textsf{Flowgen\/}} 
\newcommand{\Vincia}{\textsc{Vincia}}
\newcommand{\HTML}{\textsc{html}}

\journal{Computer Physics Communications}

\begin{document}

\begin{frontmatter}



\title{
{\normalsize \noindent
IPhT--T14/054
}\\
$\null$ \\
Flowgen: Flowchart-Based Documentation for C++ Codes}


\author{David A. Kosower}

\author{J.J. Lopez-Villarejo \corref{cor}}

\cortext[cor]{Corresponding author. \textit{E-mail address:} \texttt{jjlopezvillarejo@gmail.com}}

\address{Institut de Physique Th\'eorique, CEA--Saclay, F--91191 Gif-sur-Yvette cedex, France}

\begin{abstract}
We present the \Flowgen{} tool, which generates flowcharts from annotated C++ source code.
The tool generates a set of interconnected high-level UML activity diagrams, one
for each function or method in the C++ sources.  
It provides a simple and visual overview of complex implementations of numerical algorithms.
\Flowgen{} is complementary to the widely-used
\textsf{Doxygen} documentation tool.  
The ultimate aim is to render complex C++ computer
codes accessible, and to enhance collaboration between programmers and
algorithm or science specialists.
We describe the tool and a proof-of-concept application to
the 
\Vincia{} plug-in for simulating collisions at CERN's Large Hadron Collider.
\end{abstract}


\end{frontmatter}



{\bf PROGRAM SUMMARY}

\begin{small}
\noindent
{\em Authors:}                                                \\
{\em Program Title:}   \textsf{Flowgen}                                        \\
{\em Journal Reference:}                                      \\
{\em Catalogue identifier:}                                   \\
{\em Licensing provisions:}   GPLv2                               \\
{\em Programming language:} Python 3       \\
{\em Operating system:} Linux, MacOS, Windows                   \\
{\em RAM:} varying                                        
{\em Classification:}                                         \\
{\em External routines/libraries:}  \textsf{LibClang}, \textsf{PlantUML}    \\
{\em Nature of problem:}  To document visually the dynamic behavior of complex scientific algorithms coded in C++   \\
{\em Solution method:} Generation of a set of interconnected UML activity diagrams from annotated C++ sources\\
   \\
   \\
   \\
   \\
   \\
   \\
   \\

\end{small}

\section{Introduction}

Modern high-performance scientific computing --- such as numerical
simulations, model fitting and data analysis, and computational optimization
--- often involves fairly complex algorithms written in C++. This
complexity comes, roughly speaking, from the difference between two aspects: the high-level
(abstract) design on one hand, and the details of the specific code
implementation on the other. Moreover, the high-level design is not always easily understood from the
implementation. The intricate mixture of the two aspects creates an understanding
gap between different (groups of) developers with different expertise, hampering their
collaboration on a given code. Code developers would thus like to disentangle the
two. Software documentation for developers does not however offer a
human-understandable, high-level overview of what the code actually
\emph{does}. It also fails to keep this overview up to date with the
code. Good coding standards and strategies such as code modularity or
incremental development surely aid collaborative work, but cannot
substitute for a higher-level view.  Providing such a view in 
visual form is the challenge
we seek to meet.

\begin{figure}[!h]
    \centering
    \includegraphics[width=0.7\textwidth]{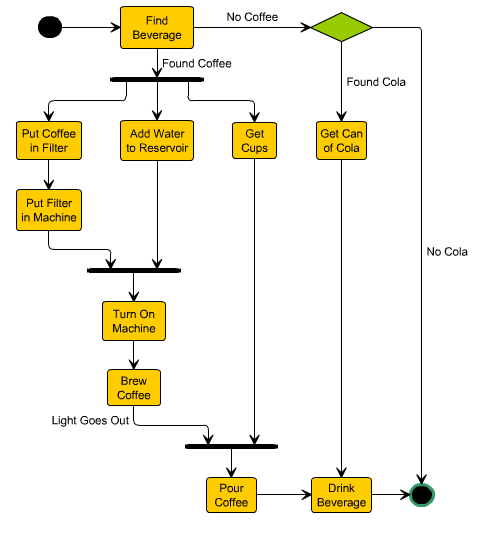}
    \caption{A UML activity diagram for thirsty people. © Object Management Group~\cite{OMG}; the 
colored version shown here is © yWorks~\cite{yWorks}.}
    \label{fig:sample_activity_diagram}
\end{figure}

Comments provide the building blocks for a successful resolution of this
challenge.  A requirement for proper comments is a part of any coding
standard nowadays.  In particular, most code includes (some) comments on the
actions carried out in the succeeding lines.  We propose using these
comments, extending them with annotations that allow us to render them along
with the code in a graphical manner.  The tool we construct uses the
annotations, along with information derived from C++ control structures, to
produce a so-called Unified Modeling Language (UML)~\cite{UML_Book} activity
diagram or \emph{flowchart} in non-specialized language (see
fig.~\ref{fig:sample_activity_diagram}).  The tool then produces a graphical
representation of this activity diagram.  This approach can be applied to the
full set of functions in a code package, or to a subset of them, as well as
to class methods (member functions).  Our focus is on producing a
\emph{``high-level''} activity diagram, one related closely to the algorithm as
designed and written by its architects and developers, rather than a
``low-level'' one more closely tied to source code.  Existing tools, described
below, generate diagrams of the latter type.

We call the tool \textsf{Flowgen\/}.  It generates a set of
interconnected high-level UML activity diagrams, one for each
annotated function or method in the C++ sources.
\Flowgen's approach is independent of any particular programming paradigm
\cite{programmingparadigms}. Its approach is modeled on that of \Doxygen{}
\cite{Doxygen}, the de facto standard tool for generating documentation
from annotated C++ sources.  It binds source code and activity diagrams
together, so that it is easier to maintain consistency between the two.  It
provides \emph{behavioral} diagrams, which complement \Doxygen's
\emph{structural} information. 
We are currently applying \Flowgen{} to the
\Vincia{} code (\url{http://vincia.hepforge.org}), a plug-in to the
high-energy physics event generator PYTHIA~8, used to simulate
proton-proton collisions at CERN's Large Hadron Collider.  We believe it
will be useful in a broad range of scientific computing codes.

\Flowgen{} seeks to provide an easy-to-read, shared standard of
communication (the activity diagrams) in a project, and thereby
to promote the goals of:
\begin{itemize}
\item Facilitating code development in international collaborations, among
  developers with different levels of expertise and coding skills, possibly 
  located in distant locations.  This involves,
 \begin{itemize}
 \item speeding up the training of new developers;
 \item allowing the participation of ``pure specialists" in the subject at
   hand, scientists with limited programming skills but an understanding
   of the high-level algorithm(s);
 \item simplifying the work of ``pure'' programmers, allowing them to 
concentrate on questions of performance;
\end{itemize}
\item Providing better \emph{backwards traceability}: a way to check that
  the final code package meets the scientific requirements;
\item Enabling iterative and incremental development of complex algorithms, 
a form of \emph{agile software development}~\cite{Agile};
\end{itemize}

In this article, we present a proof-of-concept for these goals and address
an additional one,
 \begin{itemize}
\item Increasing readability and transparency, displaying the \emph{flow of
  actions at a single glance.}
\end{itemize}

\bigskip

\Doxygen{} has emerged as a {\it de facto\/} standard for C++ \emph{structural}
documentation.  It can generate either on-line
documentation in \HTML{} format or an off-line reference manual in
$\mbox{\LaTeX}$ (or both) from a set of source files. 
In combination with the
visualization tool Graphviz~\cite{Graphviz}, it can generate class inheritance
and call graphs (basically \emph{UML class diagrams}). 
They contain structural information on how classes
relate to each other, what class members there are, and
(optionally) comments on what each class member's role is.
Annotations in the source code allow the programmer to enrich
the documentation it produces.
\Doxygen{} makes it easy to keep the documentation
consistent with the source code.  

UML was developed by the Object Management
Group~\cite{OMG}, 
and has become an industry standard.  It is intended
to help specify, visualize, and document models of software systems
using various types of diagrams.  We provide an overview of UML in~\ref{sec:UML}. 
For our purposes, there are two
main categories of diagrams: \emph{behavioral} and \emph{structural}. 
The
class diagrams that \Doxygen{} generates are structural. The
activity
diagrams~\cite{ActivityDiagrams}
we seek to generate are behavioral.  They are the 
object-oriented equivalent of flowcharts and
data-flow diagrams.  UML is used within the \emph{modeling\/} approach
to documentation, widely used in industry (though rarely
in the scientific domain).  In this approach, software applications
are designed before coding, allowing designers to work at a higher
level of abstraction.  Details can be hidden or masked, and one can
focus on different levels or aspects of a prototype.

We adopt \Doxygen's
philosophy of working with source files; \Flowgen{} produces
behavioral,
high-level UML activity diagrams as a complement to \Doxygen's structural
ones.   They are intended to describe the semantics of what a code does,
abstracted from C++ language-specific implementation
details.  
They can cover these semantics at different levels of detail,
at broad strokes corresponding to functions at the core of a call graph,
or at a finer level corresponding to leaves of a call graph.
They can also cover different levels of detail: a coarse level corresponding to
long sequences of actions accomplishing a major task, as well as zooming in 
to a single action accomplishing an elementary task.
\Flowgen{}'s complementarity to \Doxygen's
makes possible a future integration of the two tools.

A number of existing tools (both open-source and
proprietary) allow programmers to generate activity diagrams from C++ source code.  
These include: \textsf{Moritz}
(an extension to \Doxygen), \textsf{IBM Rational Rhapsody}, 
\textsf{Crystal FLOW}, \textsf{AthTek Code to FlowChart Converter}, \textsf{Code Visual to Flowchart}, \textsf{AutoFlowchart},
\textsf{devFlowcharter}.  
These tools generate diagrams based on the code, rather
than on developers' comments. 
The diagrams they produce are closely tied
to the code and are thus low-level activity diagrams in our language.

\bigskip

In following sections, we discuss different aspects of \Flowgen{} in
more detail.  In sect.~\ref{sec:a_simple_example}, we present
a simple example of annotations and the resulting output.
In sect.~\ref{sec:annotations_language}, we describe the 
code annotations used by \Flowgen; in sect.~\ref{sec:implementation},
we describe how \Flowgen{} is implemented; and in sect.~\ref{sec:tests}, we discuss
tests and lessons.  We give some concluding remarks and
an outlook in sect.~\ref{sec:conclusions}.  Because
 we have relied in general principles for its design, we believe that 
\Flowgen{} can be used for general scientific computing packages.

\section{A simple example}
\label{sec:a_simple_example}

As an example of using \Flowgen, consider a simple set of annotated C++
source files: {\tt main.cpp\/}, {\tt aux.h\/}, and {\tt aux.cpp\/}.
They are shown in the following listings,

\begin{lstlisting}[basicstyle=\ttfamily,keepspaces=true,columns=fullflexible,escapeinside={(*}{*)},caption={\tt main.cpp\/}, label=lst:main_cpp]
#include "aux.h"
#include <iostream>

int main()
{       
    int control_flag=0;
    //$ ask user whether to proceed
    std::cin >> control_flag;
    
    if (control_flag==1){
        //$ call shower
        // pointer to the object VINCIA
        VINCIA* vinciaOBJ = new VINCIA();
        vinciaOBJ->shower();  //$   
    }
    return 0;
}
\end{lstlisting}

\begin{lstlisting}[basicstyle=\ttfamily,keepspaces=true,columns=fullflexible,escapeinside={(*}{*)},caption={\tt aux.h\/}, label=lst:aux_h]
class VINCIA {
public:
    void shower();
};
\end{lstlisting}

\begin{lstlisting}[basicstyle=\ttfamily,keepspaces=true,columns=fullflexible,escapeinside={(*}{*)},caption={\tt aux.cpp\/}, label=lst:aux_cpp]
#include "aux.h"
#include <iostream>

void VINCIA::shower(){
//$ do VINCIA parton shower
std::cout << "the parton shower code would go here";
//$1 1) prepare system of partons

//$1 2) do phase 1 of shower

//$1 3)...

return;
};
\end{lstlisting}

The comments marked with \texttt{//\$} are \Flowgen{} annotations, which we
shall describe in the next section.  The tool uses them, along with
extracted knowledge of the program's control flow --- decision points (if
statements), loops, calls  --- to generate a single flowchart for each function
or method.  In our example, the tool is invoked via the following
command lines:

\begin{lstlisting}[basicstyle=\ttfamily,keepspaces=true,columns=fullflexible,escapeinside={(*}{*)},caption=command-line instructions for running \Flowgen,label=lst:runningFlowgen]
> python3 build_db.py main.cpp
> python3 build_db.py src/aux.cpp
> python3 makeflows.py main.cpp
> python3 makeflows.py src/aux.cpp 
> java -jar plantuml.jar  flowdoc/aux_files/ *.txt
> python3 makehtml.py main.cpp
> python3 makehtml.py src/aux.cpp
\end{lstlisting}

\begin{figure}[!h]
    \centering
    \includegraphics[width=0.6\textwidth]{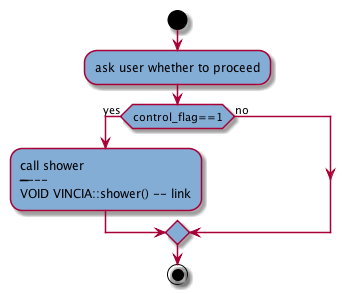}\\
    \rule{8cm}{0.4pt} \\
    \includegraphics[width=0.25\textwidth]{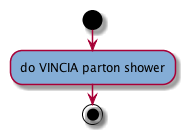}
    \includegraphics[width=0.25\textwidth]{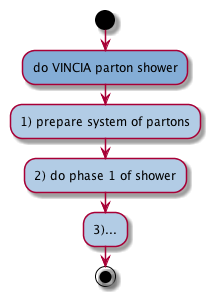}
    \caption{Output for the simple example of listings \ref{lst:main_cpp}--\ref{lst:aux_cpp}. Above: diagram for \texttt{main()} in \texttt{main.html}. Below: diagrams for \texttt{VINCIA::shower()} in \texttt{aux.html}, corresponding to zoom levels 0 (left) and 1 (right).}
    \label{fig:simple_example_diagrams}
\end{figure}

\Flowgen{} reads the source files of the project one by one and produces a set
of interrelated .html files, connected via hyperlinks, which are stored in
the folder \texttt{flowdoc/}. For this simple example,
the output consists of the diagrams in fig.~\ref{fig:simple_example_diagrams}, 
which are included in the .html files.
For code built using the \textsf{make} utility~\cite{MakeUtility}, it is easy to
adapt the makefile to run \Flowgen.

\FloatBarrier

From the \texttt{main.cpp} source file, \Flowgen{} generates
\texttt{main.html}, containing a single diagram for the function
\texttt{main()}, shown in the top diagram in
fig.~\ref{fig:simple_example_diagrams}.  \emph{Actions} are the building
blocks of the diagrams.  They correspond to sets of statements in the code
preceded by annotated lines, indicated by a leading \texttt{//\$}. The
\texttt{if} statement control structure, with condition
\texttt{control\_flag==1}, is picked up automatically and the flow paths are
displayed in the diagram. An annotation using the \texttt{//\$} prefix at the
end of a line of code serves to \emph{highlight} the call to the function or
method present on that line.  In the example, the call to the method
\texttt{VINCIA::shower()}, for which a separate diagram exists, is shown
within the action begun two lines earlier.  In addition, \Flowgen{} places a
hyperlink that allows the user to navigate to that highlighted method's
diagram present in the \texttt{aux.html} file.

\Flowgen{} generates the \texttt{aux.html} file from \texttt{aux.cpp}.
There is again a single diagram for the 
\texttt{VINCIA::shower()} method; but here, with two different \emph{zoom}
levels (0 and 1), shown in the pair of diagrams at the bottom of
fig.~\ref{fig:simple_example_diagrams}.  These zoom levels correspond
to the numerical qualifiers following the \texttt{//\$} annotations
in listing~\ref{lst:aux_cpp} (no qualifier corresponds to `$0$').

\begin{figure}[!h]
    \centering
    \includegraphics[width=0.6\textwidth]{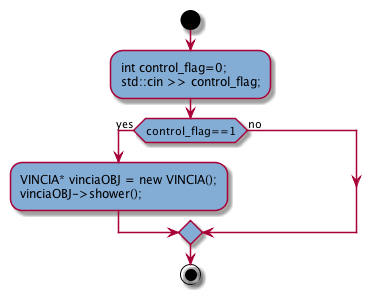}\\
    \caption{A low-level activity diagram for \texttt{main()} in listing
      \ref{lst:main_cpp}, to be contrasted with \Flowgen's high-level
      output in figure \ref{fig:simple_example_diagrams}.}
    \label{fig:lowlevel_examplediagram}
\end{figure}
\FloatBarrier

In fig.~\ref{fig:lowlevel_examplediagram}, we give an example of a
low-level activity diagram, for the \texttt{main()} function, that a tool
might generate from code lacking annotations. This is roughly the kind of
output generated by the tools mentioned in the introduction.

\section{Code Annotations}
\label{sec:annotations_language}

\Flowgen{} produces high-level UML activity diagrams, which we'll call
simply activity diagrams, 
from annotated C++ code.  It outputs these
to a set of \HTML{} files, one for each source code file.

The basic building blocks of an activity diagram are \emph{actions}, each a
statement or sequence of statements in the code.  Each action conceptually
accomplishes a discrete task.  A sequence of actions builds up an
\emph{activity}.  An activity may include different flow paths.  An
activity has a beginning and an end.  In the diagrams produced by
\Flowgen{}, these are indicated by special round symbols (see the
lower-left example in fig.~\ref{fig:simple_example_diagrams}). 
Conditional branches are indicated by diamond-shaped elements. 
The diagrams generated by \Flowgen{} are interactive. 
In particular, they allow
\emph{zooming} and \emph{browsing}. 
By zooming we mean the
possibility of inspecting the graphical description at different levels
of detail, as previously annotated by the programmer.
By browsing we mean the possibility of
navigating through the network of interconnected activity diagrams
associated with different functions or methods in a package.
Navigation is implemented using standard \HTML{} hyperlinks.

\begin{figure}[!h]
    \centering
      \begin{subfigure}[b]{0.4\textwidth}
{\ttfamily\footnotesize \begin{verbatim}
using namespace std;

int main() { 
//$ print "Hello World"
cout<<‘Hello World’;
return 0;
}


\end{verbatim} 
} 
        \end{subfigure}%
        ~ 
        \begin{subfigure}[b]{0.4\textwidth}
      \includegraphics[width=0.75\textwidth]{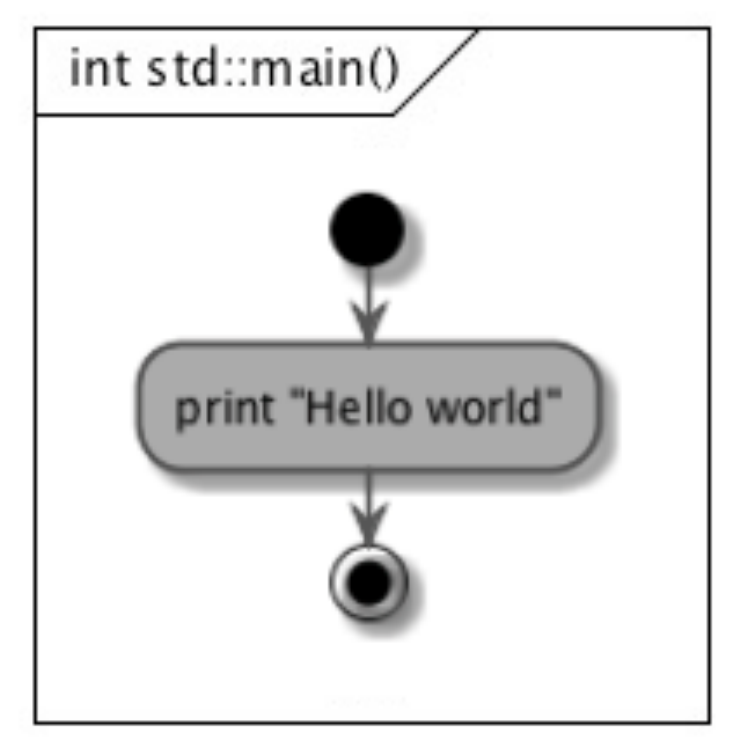}
        \end{subfigure}     
         \caption[width=0.4\textwidth]{Example of an activity: annotated code and graphical form in an activity diagram.}
    \label{fig:activity}
\end{figure}
\FloatBarrier

In the code, activities are annotated functions or methods, for example the
 C++ \texttt{main()} function (see also fig.~\ref{fig:activity}).
  (\Flowgen's annotation grammar recognizes Doxygen annotations of functions
 or methods. This feature will eventually allow their use as additional
 comments in \Flowgen{} activity diagrams.)  The actions, along with the
 level of detail to which they correspond (zoom level), are specified in the
 source code by the programmer.  The basic syntax is as
 follows,\\ \texttt{//\$ \textrm{$\langle$options$\rangle$}}
 \textsl{action description} \\ The beginning and end of the full
 activity to which the action belongs are determined by the code itself, as
 are the different flow paths within the activity.

\bigskip

An up-to-date specification for the annotation can be found on
the project's website, \url{http://jlopezvi.github.io/Flowgen}. These
include the formal specifications (Extended Backus–Naur Form). Here we
summarize the essentials:

\begin{figure}[!h]
    \centering
      \begin{subfigure}[b]{0.4\textwidth}
{\ttfamily\footnotesize \begin{verbatim}
int class::activity_method(){

int a; 
//$ do something
// we print using std::cout
std::cout << "do 1"<< endl;

//$ do other thing
std::cout << "do 2"<< endl;

return 0;
}
\end{verbatim} 
} 
        \end{subfigure}%
        ~ 
        \begin{subfigure}[b]{0.4\textwidth}
        \includegraphics[width=0.55\textwidth]{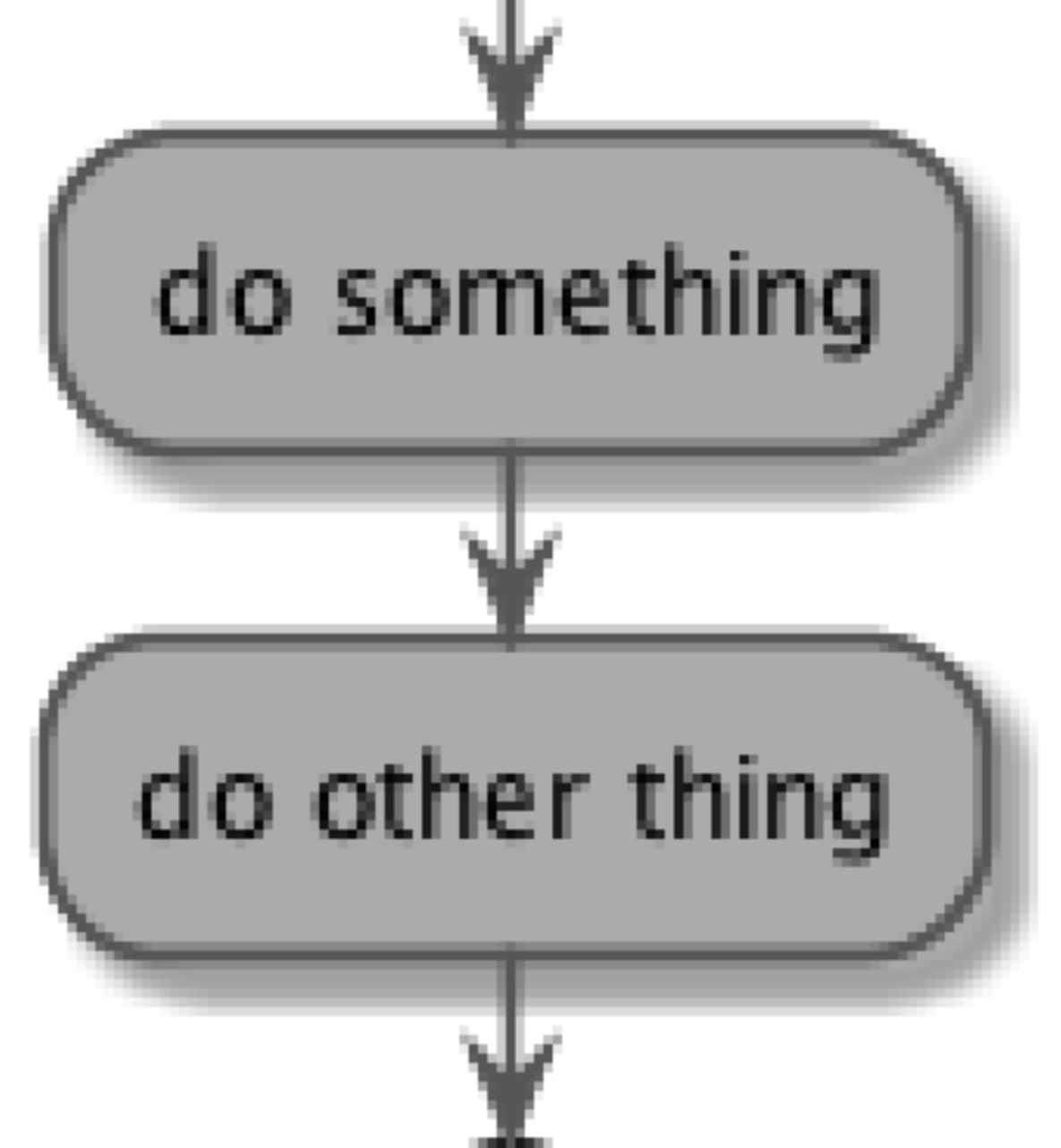}\\
        \end{subfigure}    
    \caption[width=0.4\textwidth]{Example of actions: annotated code and corresponding graphical form in an activity diagram.}
    \label{fig:action}
\end{figure}
\FloatBarrier

\begin{itemize}

\item Annotations describing actions, using the syntax given above, are the
  key added elements that allow \Flowgen{} to generate a rich description.
  (See the example in fig.~\ref{fig:action}.)  An annotation specifies what
  succeeding lines of code (up to the next annotation or an annotated
  flow-control structure) are doing.  The added `\$' distinguishes an
  annotation from a regular C++ comment, allowing the programmer to choose
  explicitly what appears in the activity diagrams.

\end{itemize}

\begin{figure}[h!]
    \centering
      \begin{subfigure}[b]{0.35\textwidth}
{\ttfamily\footnotesize \begin{verbatim}
using namespace std;

void activity_function(int a){
int c=2;
if(a>0) {
   //$ action 1
   cout<<"do 1"<< endl;
   //$ [subcondition for true]
   if (a>c) 
   {
//$ action 4
     cout<<"do 4"<< endl;
    }
}
//$ [subcondition for false]
else if(a==-1) {
//$ action 3
    cout<<"do 3"<< endl;
}
else {
cout<<"do nothing"<< endl;
}
return; 
}
\end{verbatim} 
} 
        \end{subfigure}%
        ~ 
        \begin{subfigure}[b]{0.7\textwidth}
      \includegraphics[width=1.\textwidth]{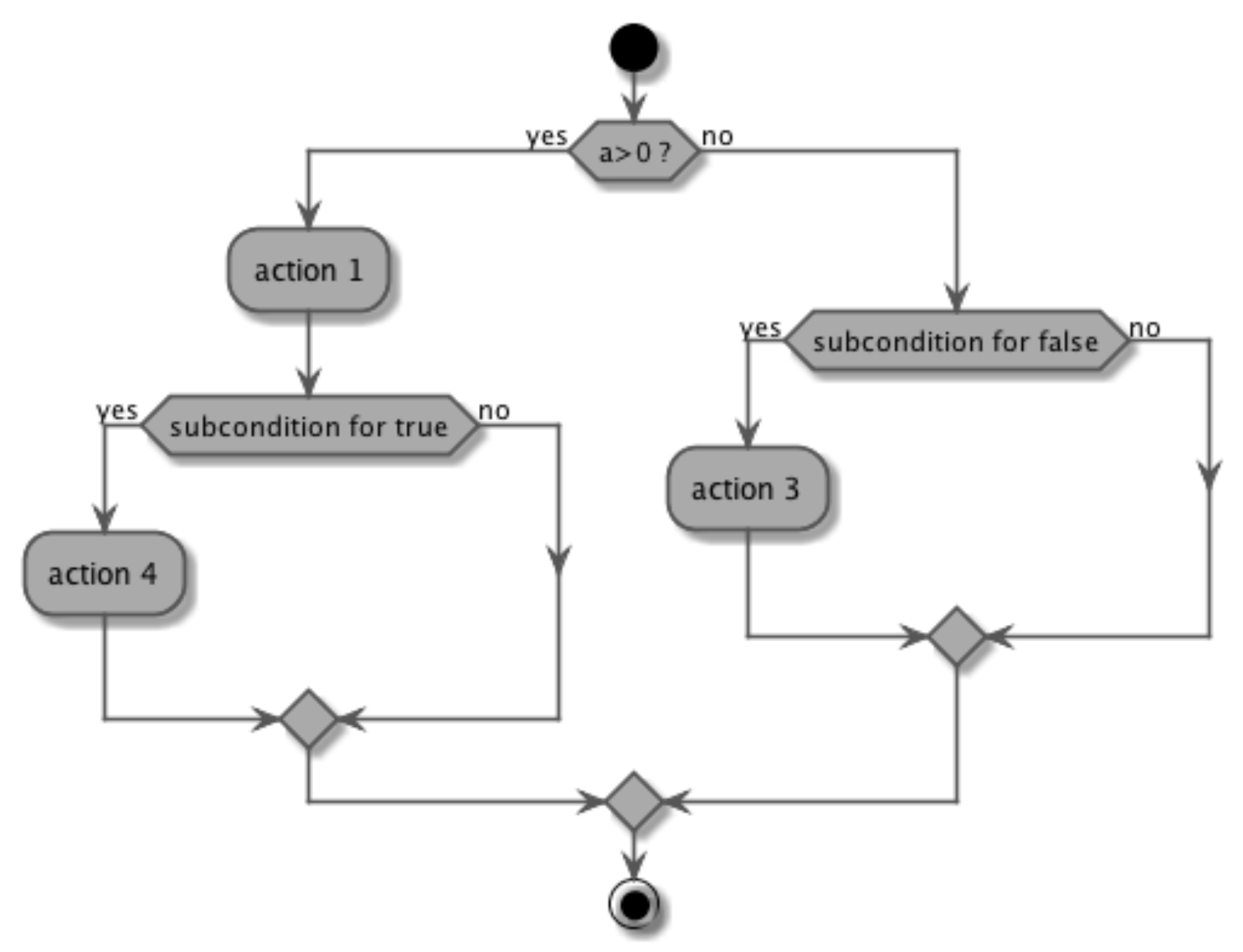}
        \end{subfigure}    

    \caption[width=0.5\textwidth]{Example of nested \texttt{if}-statements: annotated code and graphical form
in an activity diagram.}
    \label{fig:if}
\end{figure}
\FloatBarrier

\begin{itemize}

\item In \texttt{if-elseif-else} statements (see fig.~\ref{fig:if})),
  annotation allows the controlling condition to be described in a
  human-readable way. The annotation is \\ 
\texttt{//\$ }\textsl{condition description}\\ 
  which should be placed on the line immediately preceding the
  \texttt{if}, \texttt{elseif} or \texttt{else} statement that it
  describes. Loop control structures (\texttt{while, do-while, for}) 
  allow similar annotations.
\end{itemize}

\begin{figure}[h!]
    \centering    
      \begin{subfigure}[b]{0.4\textwidth}
{\ttfamily\footnotesize \begin{verbatim}
...
//$ last action
        
CODE
        
//$ [return value] 
return xVar;
... \end{verbatim} 
} 
        \end{subfigure}%
        \qquad 
        \begin{subfigure}[b]{0.4\textwidth}
           \includegraphics[width=0.6\textwidth]{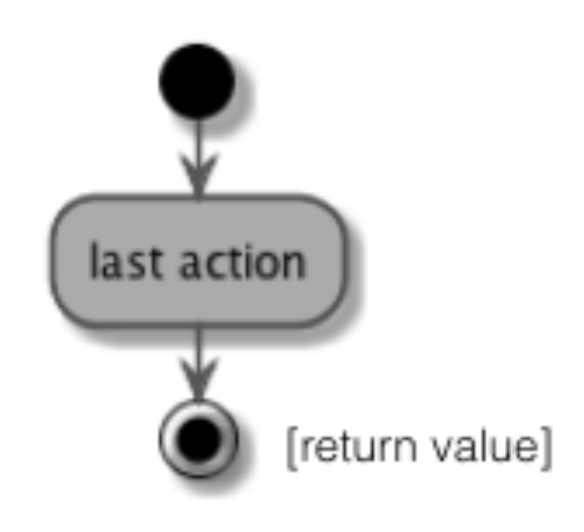}
        \end{subfigure}
\caption[width=0.4\textwidth]{Example of a \texttt{return} statement: annotated code and graphical form
in an activity diagram.}
    \label{fig:return}
\end{figure}
\FloatBarrier

\begin{itemize}

\item Annotations preceding \texttt{return} statements (see
  fig.~\ref{fig:return}) allow the return value to be specified in a
  human-readable way. (This feature has not yet been implemented.)
\end{itemize}

\begin{figure}[h!]
    \centering
          \begin{subfigure}[b]{0.4\textwidth}
{\ttfamily\footnotesize \begin{verbatim}
...
//$ <parallel> action 1
code
//$ <parallel> action 2
code
//$ <parallel> action 3
code
...
\end{verbatim} 
} 
        \end{subfigure}%
        ~ 
        \begin{subfigure}[b]{0.4\textwidth}
           \includegraphics[width=1\textwidth]{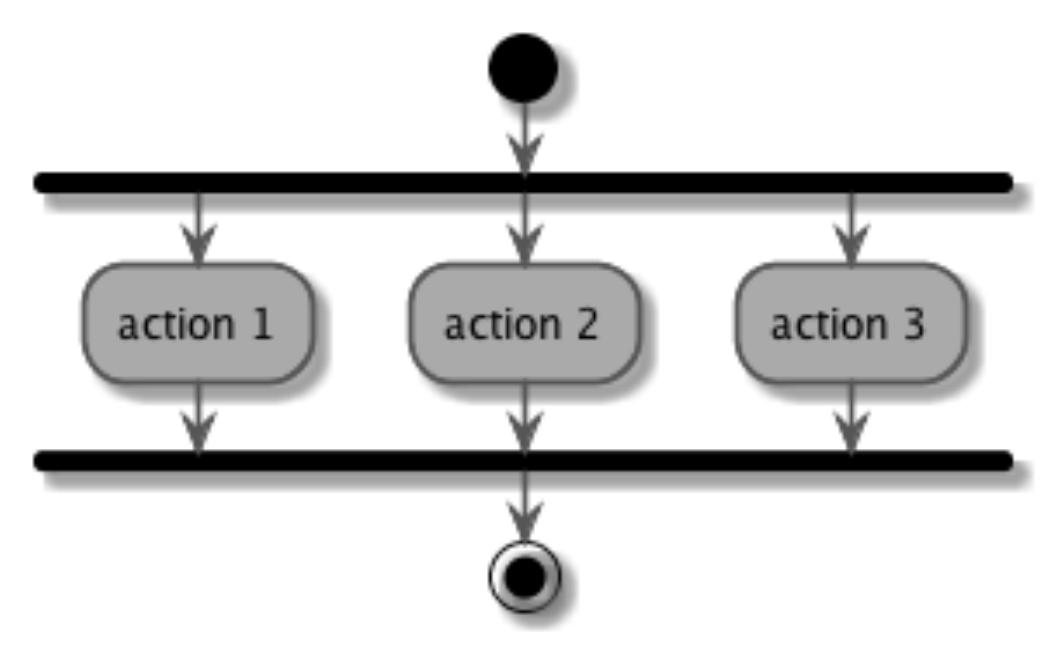}
        \end{subfigure}
    
    \caption[width=0.4\textwidth]{Example of parallel actions: annotated code and graphical form
in an activity diagram.}
    \label{fig:parallelactions}
\end{figure}
\FloatBarrier

\begin{itemize}
\item Parallel actions (see fig.~\ref{fig:parallelactions}): the tag
  `\texttt{<parallel>}' allows the programmer to indicate whether a sequence of
  actions could (in principle) be executed in parallel.  (This tag has not yet been implemented.)

\item A postfix annotation, \\ \texttt{code\_line\_with\_a\_function\_call
  //\$}\\ allows the programmer to \emph{highlight} calls to functions or
  methods (see fig.~\ref{fig:simple_example_diagrams} and the call to
  the method \texttt{VINCIA::shower()} in the example in the previous
  section).  The call will appear explicitly in the diagram.  This
  annotation also inserts a hyperlink to the diagrams for the functions or methods.  This
allows a developer to browse from the caller's diagram to the called function's
  diagram.

\item \emph{Zoom} levels: the programmer can indicate at which level of
  detail a description of an action should appear by adding an integer
  immediately after the opening `\texttt{//\$}' of an annotation (see
  fig.~\ref{fig:simple_example_diagrams} and the associated listing
  \ref{lst:aux_cpp}). Higher numbers indicate a finer level of detail; the
  zoom level is 0 by default, corresponding to the coarsest level of
  detail.  This makes different zoom levels possible in visualizing the
  \HTML{} output.
\end{itemize}

\section{Implementation}
\label{sec:implementation}

In this section we discuss the implementation of the \Flowgen{} tool.
In the first of two subsections, we discuss the requirements arising
from the specifications presented in the previous section, as well
as the choice of technologies; in the 
second, the design concept and the specific implementation.

\subsection{Requirements and Technologies}

We can classify the annotations discussed in the previous section
into three groups from a `technical' point of view. This
classification is useful to understand the requirements of our tool.
\begin{itemize}
\item[--] actions (\texttt{//\$ \textrm{$\langle$options$\rangle$}} \textsl{action description}): these will
  be given sequentially within \emph{compound statements}, that is
  sequences of code lines inside braces.  Action annotations can appear
  anywhere inside a compound statement.

\item[--] contextual annotations adding descriptions to control-flow
  structures \\(\hbox{\texttt{//\$}~\textsl{description}}): the precise position of
  these annotations is important, and they must be associated to the
  corresponding control structure (\texttt{if}, \texttt{for}, etc.)
\item[--] call highlighters (\texttt{code\_line //\$}): the preceding code
  line must be analyzed to identify function calls; and the calls should be
  matched with the corresponding diagrams for the called functions, if they
  exist.
\end{itemize}

The need to link actions to the C++ code, and in particular with detailed
knowledge of the syntax of the C++ code, makes it necessary to use
a C++ parser.  We have used the \texttt{libclang} library of the \textsf{Clang}
project
(\url{http://clang.llvm.org/}).  We use it to perform the syntax analysis
phase of C++ parsing, which yields an abstract syntax tree (AST).  This
tree is then used by \Flowgen{} to extract information it needs.

\textsf{Clang} is a C language family front-end for the LLVM
 compiler\footnote{LLVM is one of three major free C++ compilers which
 support C++11, the other two being the Intel C++ compiler and the GNU C++
 compiler (g++).}.  Clang's development is completely open-source, with
 several major software development companies involved, including Google and
 Apple. \textsf{Clang} features static analysis utilities and bindings to
 Python via a standardized library called \texttt{libclang}.  \textsf{Clang}
 also includes full support for annotations with the \Doxygen{} format
 (called ``full comments'') but, at present it leaves remaining comments out
 of the generated AST.  The documentation is at present mostly at the
 developer level.

For most of our purposes, however, regular-expression and scripting
techniques are well-suited and convenient.  For these, our language of
choice is Python (more specifically, Python 3).

\textsf{Graphviz}~\cite{Graphviz} is a standard open-source graph-drawing
package.  Beyond it, there are at least two free solutions that automate
the generation of graphs:\ \ UMLet~\ \cite{UMLet} \ \  \ \ \ \  (\url{http://www.umlet.com/})
and PlantUML (\url{http://plantuml.sourceforge.net/}).  The programs draw
diagrams from a description given in textual form in a simple and intuitive
language.  We have chosen to use PlantUML; we hope that its continued
development will also enhance the capabilities of our documentation tool.
As an interface to a visualization system, \Flowgen{} uses standard
\HTML{}, just as \Doxygen{} does. This choice allows the use of any web browser
as the visualization system and will facilitate the integration of
\Flowgen{} with \Doxygen{}.

\bigskip

We end this subsection by listing the software prerequisites for \Flowgen:
\begin{itemize}
\item LLVM-Clang 3.4 (or later) + Python3 bindings\\
\url{http://clang.llvm.org/get_started.html}\\
\url{https://github.com/kennytm/clang-cindex-python3}
\item Python3\\
\url{http://www.python.org/getit/}
\item PlantUML (included in the Flowgen distribution)\\
\url{http://plantuml.sourceforge.net/}
\end{itemize}

\subsection{Design Concept and Implementation}

\begin{figure}[h!]
    \centering
    \includegraphics[width=0.9 \textwidth]{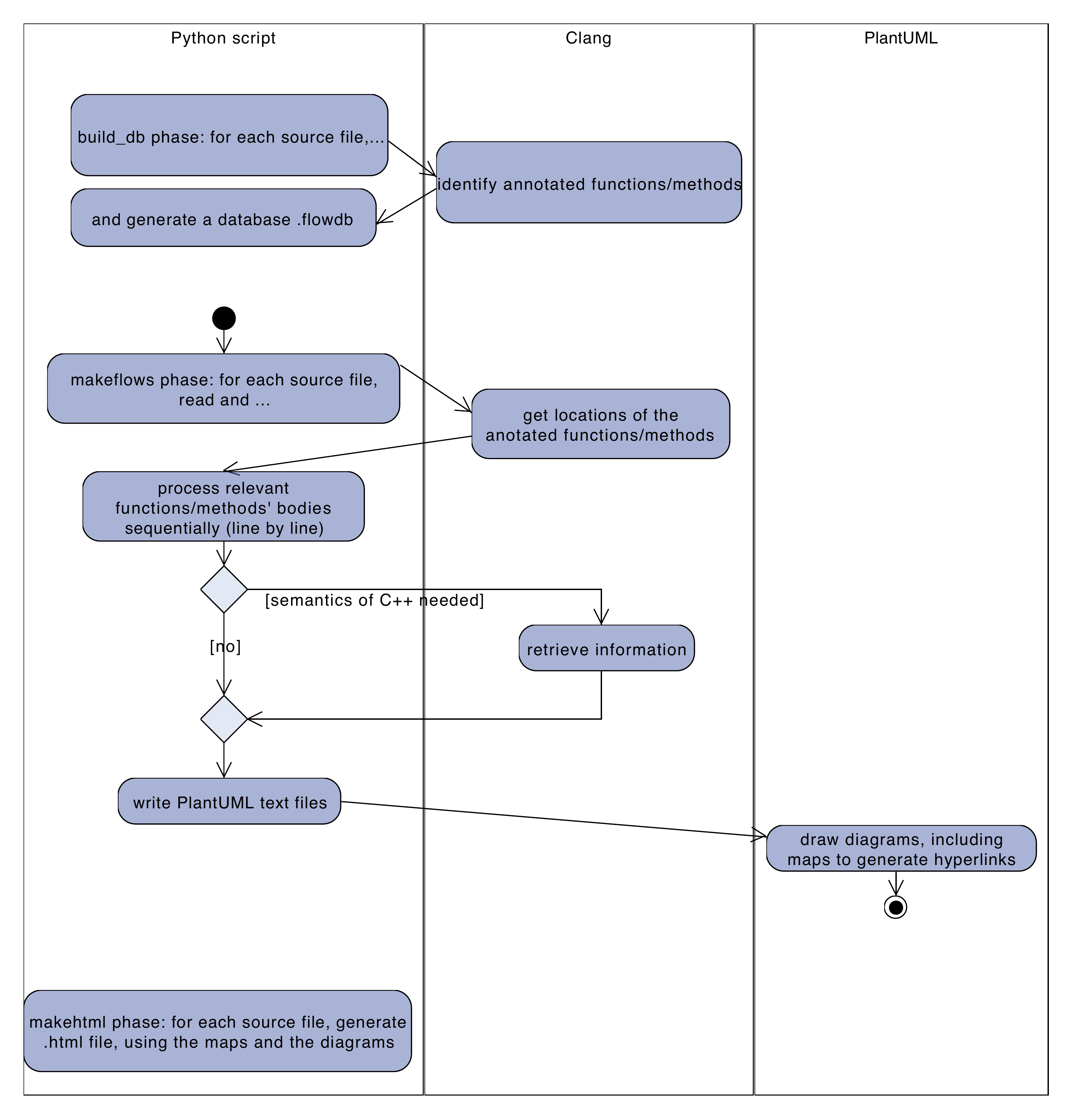}
    \caption{Flowgen's flowchart, with the three required tools and libraries: \textsf{Python}, \textsf{Clang} and \textsf{PlantUML}}
    \label{fig:flowgen_flowchart}
\end{figure}

In order to produce activity diagrams, \Flowgen{} must execute the
following steps:
\begin{enumerate}
\item Read sources
  (annotated C++ code), parse them, and link the parse tree to the
  annotations, using \textsf{Clang} and \textsf{Python3}.
\item Produce an abstract representation of
  the diagram, using \textsf{Python3}.
\item Render the abstract representation of the diagram into graphical
form, making concrete display choices for widths, lengths, fonts, colors, etc.
  This step uses \textsf{PlantUML}.
\item Embed the generated set of diagrams into \HTML{} files, to allow zooming and browsing,
  as explained in sect.~\ref{sec:annotations_language}.  This step uses
  \textsf{Python3}.
\end{enumerate}

\bigskip

We present a more detailed account of how \Flowgen{} operates in 
fig.~\ref{fig:flowgen_flowchart}. We
distinguish three phases, which however do not
precisely correspond to the list above.

In the initial `\textsf{build\_db}' phase, for each source file (headers
excluded), a database is generated which contains a list of the annotated
functions or methods. Generated database files are text files and carry the
extension `\texttt{.flowdb}'. This phase is necessary for multi-file
projects, because \textsf{Clang} cannot simultaneously process multiple
translation units.

A \textsf{Python} script controls the main phase, `\textsf{makeflows}'.  It first
reads the sources and calls \textsf{Clang} to get information on the
annotated functions or methods: namely, their starting and ending locations
in the source files.  The script then processes the corresponding ranges
line-by-line.  Some annotations (actions) are identified by simple
regular-expression parsing. More complicated structures are captured by
using the \textsf{Clang} library.  For each source file, the script writes
a corresponding text file (with suffix \texttt{.txt}) containing a
\textsf{PlantUML} description, giving the commands to draw the diagrams for
all the annotated functions or methods.  \textsf{PlantUML} is then run
(externally) in order to obtain the diagrams in \textsc{png} format, as
well as image maps in \textsc{cmapx} format. The latter are used in the
\HTML{} pages to attach hyperlinks to certain rectangular regions of the
\textsc{png} images (for example, to attach hyperlinks to calls to
functions or methods).

Finally, in the `\textsf{makehtml}' phase, another \textsf{Python}
 script generates an \HTML{} file for each source file.
The \HTML{} files include the \textsc{png} images and use the information
in the \textsc{cmapx} files.

The three phases can be automated in a makefile.

\section{Discussion}
\label{sec:tests}

We have tested our initial implementation of \Flowgen{} on a variety of
source files, which include code with nested \texttt{if} statements, calls to
functions and class methods, annotation with different zoom levels, and
links to be followed in browsing. A full, realistic
example\footnote{\Flowgen{} is not yet able to to process the
  \texttt{<parallel>} tag or the loops in this example.} for a single
method in the \Vincia{} code mentioned in the introduction can be found on
the project's website (\url{http://jlopezvi.github.io/Flowgen/}) and as accompanying files on \texttt{arXiv}. 
The example is a long procedural method where separation 
into several smaller methods is possible and may be desirable.
The code was taken from another \Vincia{} developer. 
We selected amongst those comments already present the ones that
reflected a description of the actions performed by the code,
and annotated them, including a zoom level where appropriate.  We
also indicated where parallel processing was possible.  In addition,
we annotated some conditions for \texttt{if} statements and loops.
We believe that the resulting diagram makes understanding
  the algorithm much easier, and that this understanding compensates for
the additional effort in annotation.

\bigskip

We believe that even the elementary example depicted in 
fig.~\ref{fig:simple_example_diagrams} shows the benefits of the 
tool we are proposing.  The diagrams combine two different views of
the code, a high-level semantic view on the one hand, with
code-level implementation details such as branching, or variable
and method names important to annotated activities.  It offers
a common ground for specialists of different backgrounds to collaborate
more efficiently.

We regard the present implementation as a proof of concept.  We note
in passing that the activity diagrams generated from the code by
\Flowgen{} can be modified by hand, as the \textsf{PlantUML} input
files are text files.  This could in principle be used to modify
or evolve the design of the code; the code and accompanying annotations
could then be updated to match.  The \Flowgen{} tool can thus be used
to facilitate iterative and incremental (`agile') development at a higher level
than direct coding.  We intend to apply \Flowgen{} more widely within
the \Vincia{} collaboration, and to refine it as we gain experience.
The present version of the tool is in any case available from the
project website.

\section{Conclusions and Outlook}
\label{sec:conclusions}

We have described an initial version of \Flowgen, a documentation tool that
generates high-level UML activity diagrams from annotated C++ sources.
These diagrams give a description of the dynamic behavior of the code.  The
tool is complementary to the  \Doxygen{} documentation tool, which
provides the user with structural information about static aspects of the
code.  \Flowgen{} employs annotations similar in spirit to those of
\Doxygen, designed so as not to interfere with the annotations used
by the latter.  As \Flowgen{} matures, this preserves the possibility
of combining the two tools.

A behavioral description of a software package, using activity diagrams,
allows us to see at a
glance its procedural flow of actions.
\Flowgen{} gives a graphical representation of this procedural flow,
and adds two other capabilities: 
the possibility of zooming to different levels of detail; and the 
possibility of browsing to diagrams for other called functions within
the package.

\Flowgen{} requires annotating the code to indicate the discrete actions
and select calls to hyperlink,
and optionally to add descriptions to control structures, indicate
parallelizable code, and different levels of detail for later visualization.
The additional effort to produce a basic visualization is modest;
a complete high-level description would obviously require additional
effort in rethinking the textual parts of comments.

We have designed the tool primarily for codes written in a procedural (or ``imperative'') programming paradigm \cite{programmingparadigms}, one of the paradigms possible in C++, and the
one which encompasses the bulk of scientific codes.
  It is primarily designed for developers and designers,
rather than users, but is explicitly intended to address a broad spectrum
of programming abilities, from skilled programmers to designers with
an understanding of the underlying science and algorithms but limited
programming abilities.

\medskip

\begin{figure}[h!]
\raggedleft
\includegraphics[scale=0.2]{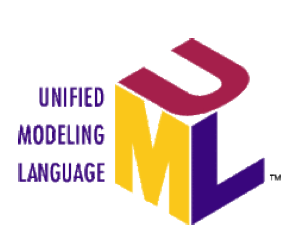} 
\includegraphics[scale=0.4]{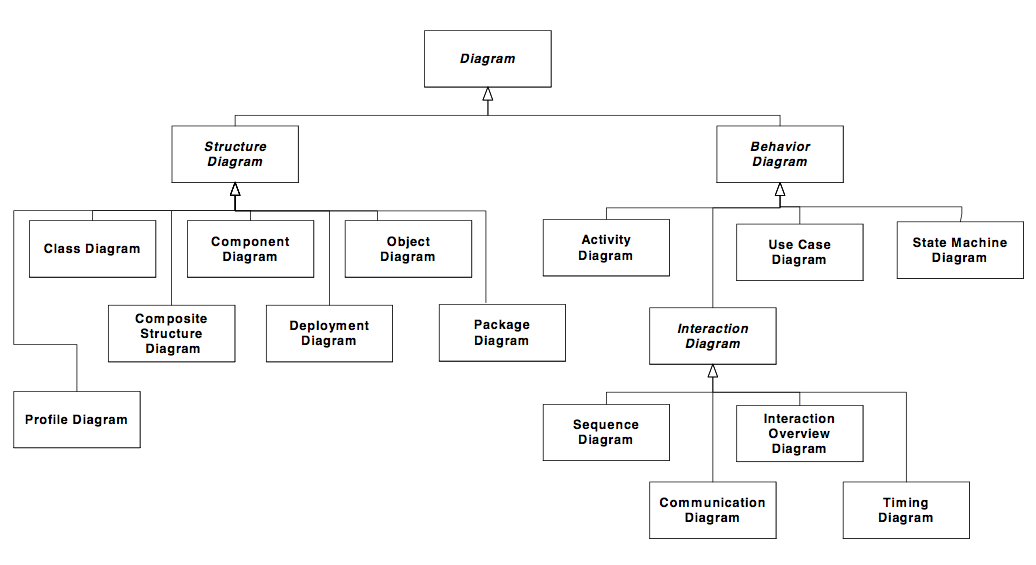}
\caption{Types of diagrams included in the UML~2.4.1 specification \cite{UML_v2.4.1}. © OMG.}
\label{fig:UMLdiagrams}
\end{figure}

\section{Acknowledgments}
We thank Serguei
Roubtsov of the Technical University of Eindhoven
for giving us feedback from a computer scientist's point of view,
and for a critical reading of the manuscript.  We also thank
Peter Skands for the original idea for \Flowgen{}, offered
during a lunchtime discussion at CERN.
This research is supported by the European Research Council under
Advanced Investigator Grant ERC–-AdG–-228301.

\appendix

\section{Unified Modeling Language --- UML}
\label{sec:UML}

The Unified Modeling Language \cite{UML_Book} is an industry standard
originally developed by the Object Management Group~\cite{OMG}.
It is intended to help specify, visualize, and document models of software
systems. It relies on object-oriented ideas such as classes and operations.
It fits most naturally with object-oriented languages and systems,
 but can be used to model other types of languages as
well. The most recent version of the UML Specification is 2.4.1 of August
2011 (\url{http://www.omg.org/spec/UML/2.4.1/}).

Fig.~\ref{fig:UMLdiagrams} shows the types of
diagrams included, organized into three
main subtypes. Seven diagram types represent static application
\emph{structure}; three represent general types of \emph{behavior};
and four represent different aspects of \emph{interactions}. 
Interaction diagrams can be considered a subtype of behavioral ones.

\end{document}